\documentclass[doublecol]{epl2}
\usepackage{amssymb}
\usepackage{graphicx}
\usepackage{color}
\usepackage{slashbox}

\institute{\inst{1} D\'{e}partement de physique and Regroupement qu\'{e}b\'{e}cois sur les
mat\'{e}riaux de pointe, Universit\'{e} de Sherbrooke, Sherbrooke Qu\'{e}bec, J1K 2R1, Canada\\
\inst{2} Institut canadien de recherches avanc\'{e}es, Universit\'{e} de
Sherbrooke, Sherbrooke, Qu\'{e}bec, J1K 2R1, Canada.}
\abstract{
Quantum critical points exist at zero temperature, yet, experimentally their influence seems to extend over a large part of the phase diagram of systems such as heavy-fermion compounds and high-temperature superconductors. Theoretically, however, it is generally not known over what range of parameters the physics is governed by the quantum critical point. We answer this question for the spin-density wave to fermi-liquid quantum critical point in the two-dimensional Hubbard model. This problem is in the $d=2,z=2$ universality class. We use the Two-Particle Self-Consistent approach, which is accurate from weak to intermediate coupling, and whose critical behavior is the same as for the self-consistent-renormalized approach of Moriya. Despite the presence of logarithmic corrections, numerical results demonstrate that quantum critical scaling for the static magnetic susceptibility can extend up to very high temperatures but that the commensurate to incommensurate crossover leads to deviations to scaling.}
\pacs{71.10.Hf}{Non-Fermi-liquid ground states, electron phase diagrams and phase transitions in model systems}
\pacs{71.10.Fd}{Lattice fermion models (Hubbard model, etc.)}
\pacs{64.60.A-}{Specific approaches applied to studies of phase transitions}
\pacs{73.43.Nq}{Quantum phase transitions}

\begin{document}

\title{Scaling and commensurate-incommensurate crossover for the $d=2$, $z=2$
quantum critical point of itinerant antiferromagnets. }
\author{S\'{e}bastien Roy\inst{1} \and A.-M.~S.~Tremblay\inst{1}\inst{2}}
\date{\today}
\maketitle

There are strong indications that quantum
critical points, i.e. critical points at zero temperature, influence the
physical properties of materials at surprisingly high temperature. But the
precise region of temperature over which this influence is felt is currently
not well understood. In solvable models of quantum critical behavior,\
\cite{Chakravarty:2005} power law scaling and universality associated with
quantum criticality were found up to temperatures of order $J/2$ where $J$ is
the exchange constant. That is in sharp contrast with classical critical
points where scaling is usually observed only in a very narrow range around
the critical point. The importance of quantum critical points \cite%
{Sachdev:2001} has thus come to the fore in the study of numerous materials,
including high-temperature superconductors and heavy-fermion materials where quantum phase transitions and power law scaling are
observed.~\cite{lohneysen:2007}

One particularly relevant case in this context is that of itinerant
electrons undergoing a paramagnetic Fermi liquid to
spin-density wave (SDW) transition in two dimensions. The Hubbard model is
the simplest microscopic model that contains this physics. There is no
analog of the Ginzburg criterion that allows us to determine the parameter
range where the influence of the quantum critical point is important. In that regime,
temperature acts like a finite-size cutoff for the correlation length $\xi $. In this paper,
we quantify the range of temperature
where quantum critical scaling is
observable in this model, in other words we find out whether details of the
Fermi surface (that lead for example to commensurate-incommensurate (C-I)
crossovers), logarithmic corrections, or interaction effects, lead to sizable deviations from quantum
critical behavior at finite temperature.

For this problem, the dynamical critical exponent $z$ is equal to two and
the corresponding universality class $\left( d+z=4\right)$ at the upper critical dimension is ill understood.
\cite{lohneysen:2007, Sachdev:2001, Pankov:2004} In particular, the standard
Hertz-Millis action for quantum critical phenomena is invalid. \cite%
{chubukov:2004, Pepin:2004} More specifically, when the
SDW is commensurate at the antiferromagnetic wave vector, it has been
suggested that all the coefficients of the Ginzburg-Landau-Wilson action
become singular and that the spin susceptibility scaling becomes $1/T^{\eta
} $ with $\eta <1.$ \cite{chubukov:2004}. The generic case where the SDW is
not commensurate should not have these singularities.

An alternative approach is the self-consistent renormalized theory of
Moriya. This theory includes logarithmic corrections.
\cite{Moriya:2006, lohneysen:2007} However, it is not adequate to
make quantitative predictions for %
deviations from quantum critical effects in the Hubbard model since it necessitates
phenomenological constants as input. In addition, it does not satisfy the Pauli principle. %
In a theory that satisfies the Pauli principle, there is an
interaction independent sum rule on spin and charge susceptibilities \cite%
{Vilk:1997} that should be enforced and, in addition, the local moment, $%
\left\langle S_{z}^{2}\right\rangle$, with $S_{z}$ the $z$ component
of the local spin, cannot exceed $\hslash^2 n/4$ when the filling $n$ satisfies  $n<1$ and $\hslash^2 (2-n)/4$ when $n>1$.
There is nothing that imposes these constraints in the theory of Moriya.

\textit{Method and model: }We use the non-perturbative Two-Particle
Self-Consistent (TPSC) approach \cite{Vilk:1997}. This approach respects the
Pauli principle, the Mermin-Wagner theorem and conversation laws. It also
contains quantum fluctuations in crossed channels that lead to Kanamori-Br%
\"{u}ckner screening. \cite{Vilk:1994}  It is valid in the
weak to intermediate coupling regime $\left( U\lesssim 6t\right) $ and not
too deep in the renormalized classical regime where a pseudogap is observed.
Numerical results
obtained from TPSC in its domain of validity
are extremely close to the numerically exact solution
obtained (barring statistical errors) with benchmark Quantum Monte
Carlo calculations on the Hubbard model. \cite{Vilk:1997, Vilk:1994,
Vilk:1995, Veilleux:1995, Moukouri:2000, Kyung:2003a}.
The approach gives a satisfactory description of the pseudogap in
electron-doped cuprates in a wide doping range. \cite{Kyung:2004,
Motoyama:2007} It has been shown to be in the $N=\infty $ universality
class, where $N$ is the number of
components in the $O(N)$ vector model. \cite{Dare:1996} Since we are
looking for deviations from universality and the theory has been benchmarked
in non-universal regimes, we argue that our results are reliable for this
question, even though we cannot claim to be completely accurate in the
$N=3$ regime. %
Nevertheless, we will demonstrate that TPSC has the same critical behavior
as Moriya theory and hence has the same logarithmic corrections. These logarithms have the same functional form as those of the
renormalization group asymptotically close to the quantum critical point, but in TPSC and in Moriya theory the mode-mode coupling term does not flow, hence the corrections may differ in the details from the renormalization group.\ \cite{lohneysen:2007}
Quantum critical behavior of the susceptibility and of the self-energy in
the closely related spin-fermion model has been discussed by Abanov et al.
\cite{Abanov:2003}

We study the $t-t^{\prime }-U$ two-dimensional Hubbard model on the square
lattice at weak to intermediate coupling,
\begin{equation}
H=-\sum_{\langle i,j\rangle ,\sigma }t_{i,j}(c_{i,\sigma }^{\dagger
}c_{j,\sigma }+h.c.)+U\sum_{i}n_{i,\uparrow }n_{i,\downarrow }
\end{equation}%
where $t_{i,j}$ are the hopping integrals, $i,j$ are the site index, $\sigma
$ is the spin label, $c_{i,\sigma }^{\dagger }$ and $c_{i,\sigma }$ are the
particle creation an annihilation operators. Each doubly occupied site costs
an energy $U$ and $n_{i,\sigma }=c_{i,\sigma }^{\dagger }c_{i,\sigma }$. The
units are such that $\hbar =1$, $k_{B}=1$ and lattice spacing is unity. All
the numerical results are presented in units where $t=1.$ The dispersion
relation is written as:
\begin{equation}
\epsilon _{\mathbf{k}}=-2t(\cos (k_{x})+\cos (k_{y}))-4t^{\prime }\cos
(k_{x})\cos (k_{y}).  \label{Dispersion}
\end{equation}

We concentrate on the behavior of the spin susceptibility. In TPSC,
the retarded spin susceptibility $\chi (%
\mathbf{q},\omega )$ is written as:
\begin{equation}
\chi (\mathbf{q},\omega )=\frac{\chi _{0}(\mathbf{q},\omega )}{1-\frac{U_{sp}%
}{2}\chi _{0}(\mathbf{q},\omega )},  \label{spin}
\end{equation}%
where $\chi _{0}(\mathbf{q},\omega )$ is the retarded Lindhard function at
wave vector $\mathbf{q}$ and angular frequency $\omega $. The effective
spin interaction $U_{sp}$ is evaluated without adjustable parameter using
the \textit{ansatz\ }\cite{Vilk:1994,Vilk:1997}%
\begin{equation}
U\langle n_{\uparrow }n_{\downarrow }\rangle =U_{sp}\langle n_{\uparrow
}\rangle \langle n_{\downarrow }\rangle \ \ (n<1)
\end{equation}%
\begin{equation}
U\langle (1-n_{\uparrow })(1-n_{\downarrow })\rangle =U_{sp}\langle (1-n_{\uparrow})\rangle \langle (1-n_{\downarrow })\rangle \ \ (n>1)  \label{ansatz}
\end{equation}%
with the local-moment sum rule that follows
from the fluctuation-dissipation theorem%
\begin{equation}
n-2\langle n_{\uparrow }n_{\downarrow }\rangle =\int_{-\infty }^{\infty }%
\frac{d\omega }{2\pi }\int_{-\infty }^{\infty }\frac{d^{2}q}{\left( 2\pi
\right) ^{2}}\frac{2}{1-e^{-\omega /T}}\chi ^{\prime \prime }(\mathbf{q}%
,\omega )  \label{sum_rule}
\end{equation}%
where $\chi ^{\prime \prime }(\mathbf{q},\omega )=\rm{Im}\chi (\mathbf{q}%
,\omega ),$ $T$ is the temperature, and $\langle n_{\uparrow }n_{\downarrow
}\rangle $ double occupancy. We dropped the site index using translational
invariance and we used the Pauli principle to write
\begin{equation}
S^{2}\equiv \langle (n_{\uparrow }-n_{\downarrow })^{2}\rangle =n-2\langle
n_{\uparrow }n_{\downarrow }\rangle .
\end{equation}

All the numerical results below are obtained using the Matsubara frequency
version of equations (\ref{Dispersion}) to (\ref{sum_rule}) without any
approximation, hence they are valid at arbitrary distance from the quantum
critical point. Before proceeding, we show however that the quantum critical
behavior of TPSC is the same as that of the self-consistent renormalized
theory of Moriya and we discuss conditions for scaling.

\textit{Analytical results near the quantum critical point. }When the
correlation length is large, one can expand the denominator of the TPSC spin
susceptibility around the wave vectors $\mathbf{q}_{i}$ where the maxima in $%
\chi _{0}$ occur to obtain

\begin{equation}
\chi ^{\prime \prime }(\mathbf{q},\omega )=\frac{2}{U_{sp}\xi _{0}^{2}}%
\sum_{i}\frac{\omega /\Gamma _{0}}{\left( \xi ^{-2}+(\mathbf{q}-\mathbf{q}%
_{i})^{2}\right) ^{2}+\left( \omega /\Gamma _{0}\right) ^{2}}.
\label{Spin_developpe}
\end{equation}%
Defining $U_{mf}=2/\chi _{0}\left( \mathbf{q}_{i}\mathbf{,}0\right) $ as the
value of the interaction at the mean-field SDW transition, the other
quantities in the previous expression are
\begin{eqnarray}
\xi ^{2} &\equiv &\xi _{0}^{2}\left( \frac{U_{sp}}{\delta U}\right) , \\
\delta U &\equiv &U_{mf}-U_{sp}, \\
\xi _{0}^{2} &\equiv &-\frac{1}{2\chi _{0}\left( 0,\mathbf{q}_{i}\right) }%
\left. \frac{\partial ^{2}\chi _{0}\left( \mathbf{q,}0\right) }{\partial
q^{2}}\right\vert _{\mathbf{q}_{i}}, \\
\frac{1}{\Gamma _{0}} &\equiv &\frac{1}{\xi _{0}^{2}\chi _{0}\left( \mathbf{q%
}_{i}\mathbf{,}0\right) }\left. \frac{\partial \chi _{0}^{R}\left( \mathbf{q}%
_{i},\omega \right) }{\partial \left( i\omega \right) }\right\vert _{\omega
=0}.  \label{Gamma_0}
\end{eqnarray}%
In the expression for the spin susceptibility, the denominators are expanded
around each of the four incommensurate wave vectors, or only around the $%
\left( \pi ,\pi \right) $ wave vector depending on the situation. We checked
explicitly that higher powers of $(\mathbf{q}-\mathbf{q}_{i})$ do not
improve the description of the C-I crossover and are not relevant.

To determine the quantum critical behavior, one subtracts the
self-consistency condition Eq.(\ref{sum_rule}) for a value of temperature
and filling close to the quantum critical point from the same equation
evaluated at that critical point
\begin{eqnarray}
S^{2}-S_{c}^{2} &=&\int_{0}^{\infty }\frac{d\omega }{\pi }\int \frac{d^{2}q}{%
\left( 2\pi \right) ^{2}}\left[ \frac{2}{\left( e^{\omega /T}-1\right) }\chi
^{\prime \prime }(\mathbf{q},\omega )\right.   \nonumber \\
&&\left. +\chi ^{\prime \prime }(\mathbf{q},\omega )-\chi _{c}^{\prime
\prime }(\mathbf{q},\omega )\right] .
\end{eqnarray}%
In the above expression $\chi _{c}^{\prime \prime }(\mathbf{q},\omega )$ is
evaluated at the quantum critical point where $\xi ^{-2}=0.$ (From now on, a subscript $c$ means that the quantity is evaluated at the quantum critical point.) One then
performs the integrals over momentum in a circular domain with cutoff $q_{B}$
and then the frequency integrals. To write the final answer, it is useful to
follow Moriya et al. \cite{Moriya:1990} and to define
\begin{equation}
T_{0}=\frac{\Gamma _{0}q_{B}^{2}}{2\pi }.
\end{equation}%
and dimensionless measures of $\xi$ and $T$%
\begin{equation}
y\equiv \frac{\xi ^{-2}}{q_{B}^{2}}\;;\;\tau \equiv \frac{T}{T_{0}}.
\end{equation}
The definition of $\Gamma _{0}$ Eq.(\ref{Gamma_0}) and the fact that $\xi
_{0}$ and $q_{B}^{-1}$ are both of the order of the lattice spacing shows
that $T_{0}$ is a temperature of the order of the Fermi energy. With these
definitions and a single maximum in the susceptibility, the self-consistency
expression takes the form%
\begin{equation}
y\left( 1-\ln y\right) =y_{0}+\frac{\tau }{\pi }\left[ \phi \left( \frac{y}{%
\tau }\right) -\phi \left( \frac{y}{\tau }+\frac{1}{\tau }\right) \right]
\label{Auto_coherent}
\end{equation}%
where terms of order $y^{2}$ have been neglected on the left-hand side. We defined
\begin{equation}
y_{0}\equiv -\frac{U_{sp}\xi _{0}^{2}}{T_{0}}\left( S^{2}-S_{c}^{2}\right)
\label{y0}
\end{equation}%
and obtained $\phi \left( x\right) $ from the second Binet log gamma
formula \cite{Erdelyi:1981}%
\begin{eqnarray}
\phi \left( x\right) &=&2\int_{0}^{\infty }dz\frac{1}{e^{2\pi z}-1}\arctan
\left( \frac{z}{x}\right) \\
&=&\ln \Gamma \left( x\right) -\left( x-\frac{1}{2}\right) \ln x+x-\frac{1}{2%
}\ln \left( 2\pi \right)
\end{eqnarray}%
with $Re[z]>0$ and $\Gamma \left( x\right) $ Euler's gamma function. The quantity $y_{0}$
in Eq.(\ref{y0}) measures the deviation from the quantum critical point.
Apart from the logarithm, the self-consistency relation Eq.(\ref%
{Auto_coherent}) has the same functional form as Eq.(2.8) in
Ref. \cite{Moriya:1990} Logarithmic corrections for that theory are mentioned without proof in Ref. \cite{Moriya:2006}.

For large local moment, $S^{2}>S_{c}^{2}$, there is an SDW ordered ground
state and $y_{0}<0.$ The case $y_{0}>0$ corresponds to the Fermi liquid
ground state and $y_{0}=0$ to the quantum critical point. The full filling
and temperature dependence of $y_{0}$ is found from the definitions of $%
U_{sp}$ and $\xi ^{-2}$. For example in the hole doped case, defining $\Delta n\equiv n-n_{c}$, we have
\begin{equation}
y_{0}=-\frac{U_{mf}\xi _{0}^{2}}{T_{0}}\frac{1}{\frac{\xi ^{-2}}{\xi
_{0}^{-2}}+1}\left[ \Delta n-\frac{U_{mf}}{2U}\frac{n^{2}}{\frac{\xi ^{-2}}{%
\xi _{0}^{-2}}+1}+\frac{U_{c,}{}_{mf}}{2U}n_{c}^{2}\right].
\end{equation}%
Thus, $y_{0}$ depends on $y=\xi
^{-2}/q_{B}^{2}$ but in the critical regime $y\ll 1$ and $y_{0}\ll 1$ so we can neglect terms of order $yy_{0}$. The quantity $y_{0}$
can then be written in the form
\begin{equation}
y_{0}=-\left( a\Delta n+bT\right)
\end{equation}%
where $a$ is a positive number.

The various limiting solutions for the critical behavior of the
dimensionless correlation length can be obtained from the self-consistency
condition Eq.(\ref{Auto_coherent}) as follows. \cite{Moriya:1990} For $%
y_{0}<0,$ one must take the limit $y\rightarrow 0$ first, then $\phi \left(
\frac{y}{\tau }\right) -\phi \left( \frac{y}{\tau }+\frac{1}{\tau }\right)
\simeq -\frac{1}{2}\ln \left( \frac{2\pi y}{\tau }\right) $ and since $y$ is
exponentially small, $y-y\ln y$ can be neglected on the left-hand side
leading to $y\simeq (\tau \exp (2\pi y_{0}/\tau ))/2\pi .$ This is the
renormalized classical regime where the correlation length grows
exponentially. At the quantum critical point $y_{0}=0$, the same limit of
the $\phi $ functions applies and one must find the solution of $-y\ln y\simeq
-\frac{\tau }{2\pi }\ln \left( \frac{2\pi y}{\tau }\right) $ which is
approximatively $y\sim \tau \ln \left( \left\vert \ln \tau \right\vert
\right) /\left\vert \ln \tau \right\vert $, as in the renormalization group. \cite{Moriya:2006} Finally, in
the Fermi liquid regime, $y_{0}>0$, the correlation length (and hence $y$) is
finite so the $\tau \rightarrow 0$ limit must be taken first and $\phi
\left( \frac{y}{\tau }\right) -\phi \left( \frac{y}{\tau }+\frac{1}{\tau }%
\right) \simeq \tau /\left( 12y\right) $ which yields $y\sim y_{0}+O\left(
\tau ^{2}\right) .$ At $\tau =0$ on the Fermi liquid side, there are
logarithmic corrections to the dependence of $y$ on $y_{0}$ asymptotically
close to the quantum critical point since $-y\ln y$ $\simeq y_{0}$ whose
approximate solution is $y\simeq -y_{0}/\ln y_{0}.$

In all regimes where $\ln y$ in the self-consistency Eq.(\ref{Auto_coherent}%
) can be neglected (large $T$) or replaced by a constant in the temperature range of
interest, one can write%
\begin{equation}
\frac{y}{\tau }\equiv F\left( \frac{\Delta n}{\tau },\frac{1}{\tau }\right)
\label{ScalingF}
\end{equation}%
where the scaling function $F$ is the solution of%
\begin{equation}
cF=\frac{y_{0}}{\tau }+\frac{1}{\pi }\left[ \phi \left( F\right) -\phi
\left( F+\frac{1}{\tau }\right) \right] .  \label{GeneralF}
\end{equation}%
with $c=1-\ln y_t$, $y_t$ being the typical value of $y$ in the range of temperature under study. We have already
discussed limiting cases of $F$ above. We demonstrate numerically below
that in the range $0.01t<T<t$ logarithmic corrections are negligible so
that scaling holds to an excellent approximation, except at the C-I
crossover.

\textit{Scaling function.}
When the explored temperature range is limited on a logarithmic scale, or when $T$ is large, logarithmic corrections can be neglected. In addition,
in the limit where $\tau $ is much smaller than $y/\tau $, the scaling
function $F$ in Eq.(\ref{ScalingF}) depends only on $\Delta n/\tau $ since
we are in the limiting case $\phi \left( F+\frac{1}{\tau }\right)
\rightarrow \phi \left( \infty \right) =0$ in the equation that defines $F,$
Eq.(\ref{GeneralF}). This case occurs when the ground state is
paramagnetic, $y_{0}>0$, or above the crossover line to the renormalized
classical regime that occurs when $y_{0}<0.$ In such cases, near anyone of the maxima located at $\mathbf{q}_{i},$ the quantity $y=\xi ^{-2}/q_{B}^{2}$ scales as
$\tau F(\Delta n/\tau,\infty )$ so the spin susceptibility Eq.(\ref{Spin_developpe}) as a function of an arbitrary scale factor $s$ obeys the scaling relation
\begin{eqnarray}
&&\chi (T,\Delta n,|\mathbf{q}-\mathbf{q}_{i}|,\omega )-R  \nonumber \\
&=&s^{\gamma /\nu }\chi _{1}(s^{1/\nu }T,s^{\phi /\nu }\Delta n,s|\mathbf{q}-%
\mathbf{q}_{i}|,s^{z}\omega ) \label{Chi_general_scaling}
\end{eqnarray}%
where the exponents have values $\gamma
=1$, $\nu =1/2,$ $z=2$ and $\phi =1$.  In the above equation, $R$ will not be important only if the incommensurate peaks are much
narrower in momentum space than the inverse correlation length$.$ Let $%
\omega =0$ for now and drop the dependence on that variable. Following the
above discussions on the behavior of the correlation length, the
susceptibility $\chi _{1}$ on the right-hand side of the last equation should
be, within log corrections, a universal function of its arguments but with the overall scale of each
argument and of $\chi _{1}$ non-universal. Setting, $%
q\equiv |\mathbf{q}-\mathbf{q}_{i}|=0$, $\omega =0$ and choosing $s$ such
that $Ts^{1/\nu }=1$ we find,
\begin{equation}
\chi (T,\Delta n,0,0)=\frac{1}{T}X\left( \frac{\Delta n}{T}\right) +R.
\label{foncquonscale}
\end{equation}%
where the scale of the function $X$ defined by this equation and an overall prefactor in front of the argument are
not universal. $X$ is the quantity we will focus on, but we note in passing
that the general form Eq.(\ref{Chi_general_scaling}) with the given
exponents implies $\omega /T$ scaling for the $q$ integrated
susceptibility.\ \cite{moi} Non-universal factors such as $U_{sp},$ $\xi _{0}$
and $\Gamma _{0}$ that enter the spin susceptibility can have some
temperature and filling dependence in TPSC that can in principle lead to
deviations to scaling. In renormalization group language, these dependencies
are the irrelevant variables whose importance we are trying to gauge to
delimit the scaling regime.

\textit{Commensurate-incommensurate crossover. }In a strict sense, the value
of $\mathbf{q}$ should be fixed at $\mathbf{q=q}_{i}\left( T=0\right) $ to
check quantum critical scaling. However, $\mathbf{q}_{i}$ itself depends
on temperature in general. At high temperature $\mathbf{q}_{i}$ equals $%
\mathbf{Q=}\left( \pi ,\pi \right) $, becoming incommensurate at low
temperature. The susceptibility there shows four symmetry related peaks for
the model we consider. \cite{Schultz:1990} The value of $\mathbf{q}%
_{i}\left( T\right) $ clearly depends on details of the Fermi surface and is
thus non-universal. The above scaling form Eq.(\ref{Chi_general_scaling})
nevertheless suggests that scaling in the $T,\Delta n$ plane as in Eq.(\ref%
{foncquonscale}) should occur when $\mathbf{q=q}_{i}\left( T\right) .$ It is
not however possible to define $\mathbf{q}_{i}\left( T\right) $ in the C-I
crossover regime. In that regime, incommensurate peaks necessarily overlap
since the second derivative of $\chi $ vanishes at $\mathbf{q=Q}$ when the
crossover begins\textbf{,} reflecting the fact that there is a broad maximum
at $\mathbf{Q}$ that is splitting into four overlapping peaks. $R$ in the
general scaling function Eq.(\ref{foncquonscale}) is not negligible in the
C-I crossover region. On general grounds then, we expect deviations to
scaling there. One may think that a better strategy to prove scaling is to
measure the correlation length $\xi $ as a function of $T$ an $\Delta n,$
but $\xi $ cannot be determined in the C-I crossover regime for the same
above reasons.

From now on, we thus look for scaling with the susceptibility evaluated at its maximum, $%
\chi (T,\Delta n,|\mathbf{q}_{\max }-\mathbf{q}_{i}|,0).$ This is a well
defined quantity experimentally and far from the C-I crossover we will have $%
\mathbf{q}_{\max }=\mathbf{q}_{i}\left( T\right) .$ %After we show the
%numerical results, we explain them analytically.

\begin{figure}[tbp]
\centering%
%TCIMACRO{\TeXButton{dn=0}{\includegraphics[width=2in]{figure3_1.eps}}}%
%BeginExpansion
\includegraphics[width=2in]{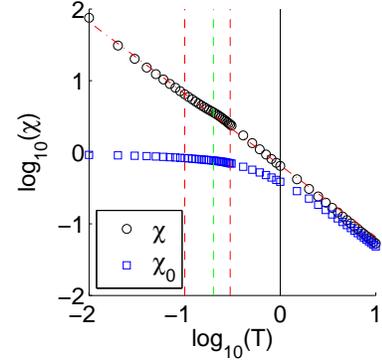}%
%EndExpansion
\caption{(Color online) $\log _{10}(\protect\chi )$ as a function of $\log _{10}\left(
T\right) $ for $U=4t$ and $t^{\prime }=0$ (open circles) at $n_{c}=0.84$ and
for the corresponding non-interacting susceptibility at $U=0$ (open
squares). The vertical dashed lines indicate the commensurate to
incommensurate crossover region. The red dash-dotted line has a slope $-1$. The
black vertical line simply indicates $T=t.$ }
\label{dn=0}
\end{figure}

\textit{Numerical results: }Let us first verify the scaling at the quantum
critical point $\Delta n=0$. Fig. (\ref{dn=0}) shows a log-log plot of both the
interacting (open circles) and noninteracting (open squares)
susceptibilities as a function of temperature for two different sets of
parameters.

For temperatures larger than hopping $t,$ one obtains trivial $1/T$ scaling
for both the interacting and non-interacting susceptibilities. While the
non-interacting susceptibility flattens at lower temperature, the
interacting susceptibility shows quantum critical $1/T$ scaling down to the
lowest temperature we could reach, namely $T=0.01t.$ We will see that the $%
1/T$ scaling at $T>1$ that comes from the non-interacting susceptibility
does not obey the scaling equation Eq.(\ref{foncquonscale}). It is also
clear from Fig. (\ref{dn=0}) that deviations to scaling occur in the C-I
crossover regime delimited by the vertical red lines. It is remarkable
however that the same straight line fits both the commensurate and the
incommensurate regimes. This suggests that non-universal scale factors are very similar on either sides of the commensurate-incommensurate transition. The slight upward curvature at the lowest temperatures is not inconsistent with effects of logarithmic corrections.

To verify the full scaling Eq.(\ref{foncquonscale}), we plot $T\chi $ as a
function of $|\Delta n|/T$ on a log-log plot in Fig. \ref%
{fig1}. We take values of $n$ on the Fermi liquid side of $n_{c}$. For a given band structure and interaction, it is only when
one has found the correct values of the critical $n=n_{c}$ that all the
curves for different fillings and temperature collapse on the same curve. We
found, when $t^{\prime }=0$, that $n_{c}=0.926,$ $0.840$ and $0.795$ for $%
U=2t,4t$ and $6t$ respectively and $n_{c}=1.180$ for the electron-doped case
with $U=6t,$ $t\,^{\prime }=-0.05t.$ More values can be found in the thesis
which is the basis for all the results of the present paper.\ \cite{moi}

\begin{figure}[tbp]
\centering\includegraphics[width=3.3in]{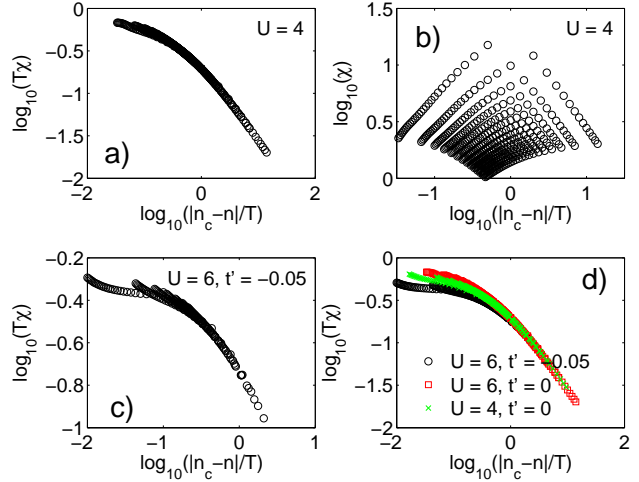}
\caption{(Color online) a) $\log _{10}(T\protect\chi )$ as a function of $%
\log _{10}\left( \frac{|\Delta n|}{T}\right) $ for $U=4t$ and $t^{\prime }=0$.
b) Shows the data of figure a) with the susceptibility unscaled $\log _{10}(%
\protect\chi )$ c) $\log _{10}(T\protect\chi )$ as a function of $\log
_{10}\left( \frac{|\Delta n|}{T}\right) $ for $U=6t$ and $t^{\prime }=-0.05t$.
d) compares the scaled data $\log _{10}(T\protect\chi )$ as a function of $%
\log _{10}\left( \frac{\Delta n}{T}\right) $ for $U=4t$ and $t^{\prime }=0$
(green crosses), $U=6t$ and $t^{\prime }=0$ (red open squares) and $U=6t$
and $t^{\prime }=-0.05t$ (black open circles). On figures a) and c) we can
see deviations from the scaling at $\log_{10}\frac{|\Delta n|}{T}<-1$ that are due to
the commensurate incommensurate crossover (see text). In these figures, for
$t^{\prime }=0$ we considered hole doping with $n_{c}-n>0$, the smallest $n$
being $n=0.7.$ For $t^{\prime }=-0.05t,$ we took electron doping with the
largest $n$ equal to $n=1.3$ and $n-n_{c}$ as argument of the log. }
\label{fig1}
\end{figure}

The straight line of slope $-1$ at large $\Delta n/T$ in Figs.\ \ref{fig1}%
a,c,d. corresponds to the Fermi liquid regime where both the susceptibility
and the correlation length are temperature independent, but diverge as one
approaches the quantum critical point. In that regime, $\chi $ scales as $%
\xi ^{2}\sim 1/\Delta n$ when logarithmic corrections are negligible. The $%
1/T$ scaling of $\chi $ corresponds to plateaus on the left of Figs.\ \ref%
{fig1}a,c,d. The deviations from a plateau come from the C-I crossover. To
show that the scaling is non-trivial, in Fig.\ \ref{fig1}b we do not
multiply the susceptibility by $T$ on the vertical axis. The lined up
circles that can be caught by the eye correspond to different temperatures
for a given filling $n,$ the fillings closest to $n_{c}$ being to the left.

Scale factors depending on band structure and interaction strength should
not influence the shape of the scaling function. A simple translation in the
$T-\Delta n$ plane of the curves for different parameters should allow all
of them to collapse. In Fig.\ \ref{fig1}d, we show scaling functions for
various parameters but without translation for non-universal factors. One
sees that if there were no deviations to scaling associated with the C-I
crossover in the plateau region, simple translation would make all the
curves nearly collapse. This also shows that logarithms do not
have a large influence on scaling in this temperature range.

\begin{figure}[tbp]
\centering\includegraphics[width=3.3in]{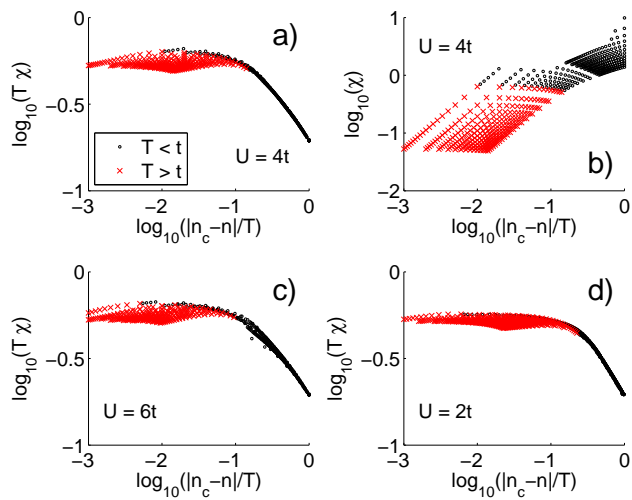}
\caption{ (Color online) The black circles are the data for $T<t$ and the red crosses for $%
T>t$. a) Scaled data $\log _{10}\left( T\protect\chi \right) $ for $%
U=4t$. b) Unscaled values $\log _{10}\left( \protect\chi \right) $ for
the same $U$. c) $\log _{10}\left( T\protect\chi \right) $ for $U=6t$
d) for $U=2t$. All four panels are at $t^{\prime }=0$. The high temperature
limit of the quantum critical scaling is of the order $T\sim t.$}
\label{fig2}
\end{figure}

In Fig. \ref{fig2}, data analogous to those in Fig.\ \ref{fig1} are
represented by black open circles and are filtered out near the C-I
crossover. The missing data is particularly clear in Fig.\ \ref{fig2}b where
we do not scale the vertical axis. If $T_{i}$ is the temperature where the
crossover occurs for a given doping, the data were filtered in the range $%
T_{i}-\Delta T<T<T_{i}+\Delta T$ ($\Delta T\sim 0.2t$) for densities $n_{c}-n<0.04$. For larger values of $n_{c}-n$, the data is sufficiently far from the C-I crossover that no filtering is required.
%and $\Delta T=0$ if $n-n_{c}>0.04$.
The remaining data are those beyond the
C-I crossover both above (commensurate) and below (incommensurate) $T_{i}$.
%
%The small descrepencies in the figures \ref{fig1} are due to the commensurate to incommensurate boundary. Since there is no simple way to express the TPSC spin susceptibility with an approximate form coherent with equation \ref{gensum}, there is no direct way to extract the actual correlation length from the spin susceptibility and we can't evaluate $\mathbf{q}_{i}$. For the scaling rule to work flawlessly, the invariant spin susceptibility must be evaluated at $\mathbf{q}=\mathbf{q}_{i}$, where $\mathbf{q}_{i}$ is not known. It is evaluated at $\mathbf{Q}_{max}$ instead which explain the deviations.
One sees that a plateau is recovered (black open circles) for all three values of the interaction
strength appearing in Figs.\ \ref{fig2}a,c,d as expected in the quantum
critical regime.

We now turn to the high temperature limit of the quantum critical scaling.
While the black open circles in Fig. \ref{fig2} are for $T<t,$ those for $%
t<T<10t$ are represented by red crosses. The deviations to scaling for $%
t<T<10t$ are obvious. Even though the non-interacting susceptibility scales
as $1/T\;$for $T>t$ as we saw in Fig.$\;$\ref{dn=0}, it does not pollute the
scaling associated purely with the quantum critical point. The latter occurs
for $T<t,$ with the caveat concerning the C-I crossover. The maximum $T$ for
scaling, $T\sim t$, is an important result that applies in the weak to
intermediate coupling regime we have considered here. Clearly
quantum critical behavior must disappear at $U=0,$ so there should be some $%
U $ dependence to the upper temperature cutoff. At the intermediate coupling
values that we considered, the temperature range over which quantum critical
scaling is observed should be compared to what would have been naively estimated
by substituting $U=4t$ and $U=6t$ in $J=4t^{2}/U$ \cite{Chakravarty:2005}, obtaining respectively $%
J/2=t/2$ and $J/2=t/3.$ Basically, the upper limit of $T\sim t$ is essentially the degeneracy temperature for Fermi-Dirac
statistics, which is of the same order as $T_{0}$. The irrelevant temperature dependencies of all quantities are thus on this scale.

\textit{Conclusion. }The quantum critical
behavior of TPSC for the $d=2,$ $z=2$ universality class is the same as that
of the self-consistent renormalized theory of Moriya, hence it includes
logarithmic corrections. In TPSC there is no adjustable parameter. By explicit numerical
calculations away from the renormalized classical regime of the $d=2$
Hubbard model in the weak to intermediate coupling, we have been able to
show that logarithmic corrections are not really apparent in the range of temperature $0.01 < T < t$ and that the maximum static spin
susceptibility in the $(T,n)$ plane obeys quantum critical scaling. However,
near the commensurate-incommensurate crossover, one finds obvious
non-universal temperature and filling dependence. Everywhere else, the $%
(T,n) $ dependence of the non-universal scale factors is relatively weak.
Strong deviations from scaling occur at temperatures of order $t$, the
degeneracy temperature, reflecting the fact that the temperature dependence
of most irrelevant terms is on the scale of the Fermi energy. That high temperature limit should be contrasted with $J/2$ found in the strong coupling case. \cite{Chakravarty:2005} In generic cases the upper limit $%
T\sim t$ is well above room temperature. In experiment however, the
non-universality due to the C-I crossover may make the identification of
quantum critical scaling difficult. And since the $(T,n)\mathbf{\ }$dependence
of $\mathbf{q}_{i}$ is non universal, one may encounter cases where this is
in practice impossible.

Electron-doped high-temperature
superconductors appear as an ideal system to check quantum critical scaling
since they seem well described by the $d=2$ one-band Hubbard model at weak
to intermediate coupling. \cite{Kyung:2004, Motoyama:2007} And experiments
\cite{greene1,Charpentier:2008} strongly suggest the presence of a quantum critical point in
these materials. In the case of heavy fermions there are examples of
SDW-Fermi liquid quantum critical behavior. \cite{lohneysen:2007}
However, these are multiband systems where there are additional energy scales, such as the Kondo coherence scale, so our results would apply only in regimes where an effective one-band Hubbard model applies.

A.-M.S.T. would like to particularly thank %
A. Chubukov and J. Schmalian for informative and stimulating
comments on our work, and the Max-Planck Institute for the Physics of
Complex Systems for hospitality. We are also indebted to S. Chakravarty, and
S. Sachdev for discussions at the Aspen Center for Physics and most
importantly to B. Kyung for sharing with us his experience with TPSC on
numerous occasions. Numerical calculations were performed on RQCHP computers
and on the Elix cluster. The present work was supported by NSERC
(Canada), FQRNT (Qu\'{e}bec), CFI (Canada), CIFAR, and the Tier I Canada
Research chair Program (A.-M.S.T.).

%\bibliographystyle{eplbib}
%\bibliography{article.bib}

%\begin{thebibliography}{99}
%\end{thebibliography}

\end{document}